\documentclass[graybox]{svmult}

\usepackage{type1cm}        
\usepackage{makeidx}         
\usepackage{graphicx}        
\usepackage{multicol}        
\usepackage[bottom]{footmisc}

\usepackage{newtxtext}       %
\usepackage{newtxmath}       

\usepackage{subfig}
\usepackage{graphicx}
\usepackage{amsmath}
\usepackage[version=4]{mhchem}
\usepackage{siunitx}
\usepackage{longtable,tabularx}
\usepackage{subfiles} 
\usepackage{pdfpages} 
\usepackage{csvsimple} 
\usepackage{tikz} 
\usetikzlibrary{shapes, arrows, positioning} 

\makeindex             

\begin{document}

\title*{A model agnostic eXplainable AI based fuzzy framework for sensor constrained Aerospace maintenance applications}
\titlerunning{XAI and Fuzzy for constrained sensor data}
\author{Bharadwaj Dogga*, Anoop Sathyan, and Kelly Cohen}
\authorrunning{Dogga et al.}
\institute{Bharadwaj Dogga*, Ph.D. Student \at AI Bio Lab, Digital Futures, University of Cincinnati, Cincinnati, OH 45221, USA  \\ \email{doggabj@mail.uc.edu}
\and Anoop Sathyan, Research Associate \at AI Bio Lab, Digital Futures, University of Cincinnati, Cincinnati, OH 45221, USA \\ \email{sathyaap@ucmail.uc.edu}
\and Kelly Cohen, Director \at AI Bio Lab, Digital Futures, University of Cincinnati, Cincinnati, OH 45221, USA \\ \email{cohenky@ucmail.uc.edu}
}
%
%
\maketitle

\abstract{Machine Learning methods have extensively evolved to support industrial big data methods and their corresponding need in gas turbine maintenance and prognostics. However, most unsupervised methods need extensively labeled data to perform predictions across many dimensions. The cutting edge of small and medium applications do not necessarily maintain operational sensors and data acquisition with rising costs and diminishing profits. We propose a framework to make sensor maintenance priority decisions using a combination of SHAP, UMAP, Fuzzy C-means clustering. An aerospace jet engine dataset is used as a case study.}

\section{Introduction}
\label{sec:1}

The extensive amount of data gathering with regards to jet engines begins to take toll on the maintenance pipelines, once an engine starts to enter the final years of its life-cycle. Negligence in this domain results in extensive damage both in terms of parts of an airplane engine that is damaged due to bad maintenance \cite{de2023exploring} along with associated costs in the air travel supply chain. Sensor constraint is one proposed solution but is limited by decision making with regards to which sensors to limit maintenance from. One nominal approach for sensor constraint associated savings for end of cycle engines is to optimize the number of sensors being used for data gathering and sensor maintenance associated costs. To solve this need, we propose an Industrial AI framework to identity the most contributing sensors for an aerospace engine to help with preventive maintenance scheduling. The dataset being considered is a NASA C-MAPSS ’08 \cite{saxena2008damage} which contains sensor data from engine stations. While a more recent dataset addresses timeseries data with known failure modes, the 2008 version presents a unique opportunity due to the lack of information and opens up the design space to have literature based constraints like theoretical RUL values\cite{heimes2008recurrent}.
The initial intent for this dataset was to serve as a test bench for a data competition but since then, it has expanded as a test bed to demonstrate recent machine learning advancement in the PHM domain ranging from Graph Neural Networks\cite{li2019directed}, to LSTMs\cite{tian2023spatial}, and transformer based models\cite{zhang2023integrated}. Furthermore, systematic review papers\cite{siraskar2023reinforcement}\cite{lei2018machinery} have been written comparing the state of the art for these methods along with identifying future challenges\cite{rezaeianjouybari2020deep}.

Explainable methods have been used to augment decision making in the past for statistics \cite{pickering2024explainable}, uncertainty quantification \cite{cohen2023trust} and aerospace applications\cite{dogga2024explainable}. Methods have been expanded to include semi-supervised fault diagnostics \cite{cohen2023shapley}, for semiconductor applications\cite{ cohen2023trust} \cite{cohen2022semi} and airplane engine fault applications. Fuzzy based clustering has also been proven to be resourceful for this approach\cite{cohen2021deep}. The clusters obtained from such methods can be easily evaluated using scikit-learn\cite{scikit-learn}'s clustering performance evaluation modules.

Model agnostic eXplainable AI (XAI) method SHapley Additive exPlanations (SHAP) \cite{lundberg2017unified} is used to obtain the explanations for this model. Uniform Manifold approximation and Projection (UMAP)\cite{mcinnes2018umap} is the go to dimension reduction method to help with dimensional reduction of data and was picked over other dimension reductions due to its merit with relatively keeping density and distances intact when dealing with higher dimensional data\cite{pealat2021improved}. Advances of such approaches can be scaled to a multi-model level based on this same framework.

\section{Methodology}
\label{sec:2}

This paper demonstrates a novel explainable fuzzy c-means based framework for Engine maintenance scheduling applications using the following tools:

\begin{enumerate}
    \item SHapley Additive exPlanations (SHAP) for Shapley value analysis to identify the top variables to monitor for a given engine type
    \item Uniform Manifold approximation and Projection (UMAP) for dimensionality reduction of captured engine data
    \item Fuzzy c-means clustering for capturing RUL bins info for engines that need to schedule their maintenance
\end{enumerate}

It addresses the need for model agnostic explainable fuzzy tool that can predict when a late lifecycle aerospace engine would need to schedule maintenance based on partially labeled station data and already available history of maintenance cycles. Then use SHAP to identify a similar maintenance bins using 70\% less data to deliver reliable sensor cost savings for the engine operators. To compare the metrics, we study four cases of data from a single dataset FD001 from the aforementioned NASA C-MAPSS ’08 dataset.

\subsection{Data Pre-processing}

\label{subsec:2}
C-MAPSS ’08 Data set contains four different engine error cases ranging from FD-001 to FD-004. The sensor data measured for these data is spread across 26 columns, along with an engine ID and cycle number within each engine ID. While useful for timeseries data, engine IDs are not useful for individual cycle modelling and are hence discarded from our data. Engine cycle numbers are used both as a cycle parameter and an input to calculating remaining useful life (RUL). Though mostly experimental, previous literature points to a linear piecewise polynomial fit with a plateau peak RUL beyond the 125 to 130 range with values dropping linearly to zero after 125. A similar target RUL is used for this training and model fit to simulate a well labeled data set. This piecewise function is denoted in equation (\ref{piecewise_linear_interpolant}). Here, \emph{c} represents the cycle number of a given engine and \emph{A} is the maximum number of cycles for an engine.


\begin{equation}
\label{piecewise_linear_interpolant}
    RUL(c) =
    \begin{cases} 
    125, &  c \leq (A - 125) \\
    A - c, & (A - 125) < c < A
    \end{cases}
\end{equation}

\begin{equation}
\label{normalization_of_data}
x' = \frac{2*(x - x_{min})}{x_{max} - x_{min}} - 1, \; x' \in [-1, 1]
\end{equation}

\begin{equation}
\label{denormalization_of_data}
x = (x' + 1) \frac{x_{max} - x_{min}}{2} + x_{min}
\end{equation}

As mentioned, we used only FD-001 engine and the complete cycle data was present only for test part of the data set which had 20,631 rows of data. To avoid interference of constant data columns, only non-zero range columns were considered for this modelling which were then normalized and scaled into $[-1, 1]$ to take advantage of maximum floating points with equation (\ref{normalization_of_data}). Once the Neural Network is run, we will then de-normalize the data using equation (\ref{denormalization_of_data}).

\subsection{Neural Network}
\label{subsec:3}
\begin{figure}[hbt!]
\centering
\includegraphics[width = 0.65\linewidth]{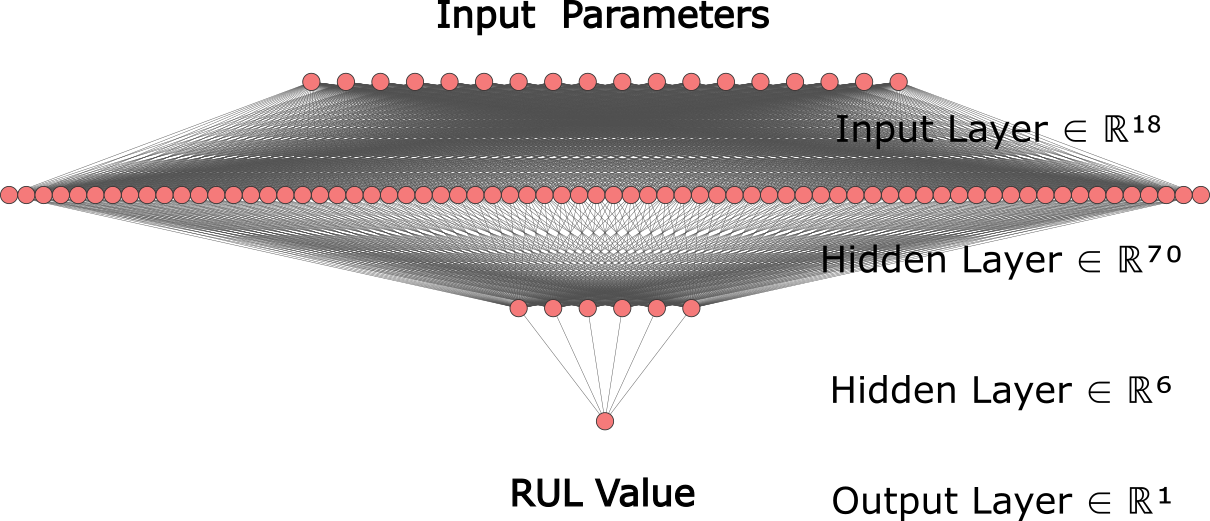}
\caption{Neural Net model: C-MAPSS '08 FD-001 dataset}
\label{fig:NN1}
\end{figure}

A regression problem approach was taken to model an artificial Neural Network for estimating RUL. A few variations of Hypermeters were also tested which resulted in two rectified linear unit(ReLU) hidden layers with 70 nodes and 6 nodes respectively. Input parameters were linear units with their number equal to the number of non-zero input units and output layer was a single ReLU node that stored RUL values. 80\% of the dataset’s 20,631 rows were used as training data and the remainder was assigned to testing data. The neural net is visualizes in Figure \ref{fig:NN1}.

\begin{equation}
\label{l2_norm_MSE}
MSE = \sum_{i=1}^{n}(x_i-\overline{x_i})^2
\end{equation}

Adam Optimizer with a learning rate of 1e-4 was used with a batch size of 32. The error function L2 norm - equation (\ref{l2_norm_MSE}) was used to quantify the performance of NN results. The resulting RULs were then assigned a \emph{maintenance bins} based on the predicted RUL value as shown in Table \ref{table:bins_assignment}. These maintenance bins will help us understand which of these machines need to be scheduled for maintenance based on the predictors.

\begin{table}[ht]
\centering
\begin{tabular}{c|c}
\hline
RUL Range  & Maintenance Bins \\
\hline
\\
125 and above & Great \\ 
125 to 75 & Good \\ 
75 to 50 & Okay \\ 
50 and below & Schedule \\
\hline
\end{tabular}
\caption{Assigned RUL maintenance bins}
\label{table:bins_assignment}
\end{table}

\subsection{eXplainable AI and SHAP}
eXplainable Artificial Intelligence (XAI) tool SHAP module calculates the Shapley value for a feature and sample \emph{i} as shown below in equation (\ref{SHAP_equation}). \emph{F} is the total number of features, and \emph{S} is a subset on which we are training the model. One of the most important features of Shapley values is local accuracy which is defined by equation (\ref{local_accuracy}) and denotes that the sum of all the contributions sums the prediction of the result.

\begin{equation}
\label{SHAP_equation}
    \phi_i=\sum_{S\subseteq F\setminus\{i\}}\frac{|S|!(|F|-|S|-1)!}{|F|!}\left(f_{S\cup\{i\}}(x_{S\cup\{i\}})-f_s(x_S)\right)\
\end{equation}

\begin{equation}
\label{local_accuracy}
    g(z')=\phi_0+\sum_{j=1}^M\phi_jz_j'
\end{equation}

In our framework, we will be using this property of  SHAP to capture top five variables that effect the RUL prediction the most and then contrast them with the RUL bins defined from all the other variables.

\subsection{Cluster Analysis and Verification}
\label{subsec:4}
UMAP is used for dimension reduction. The eighteen dimensions of data are reduced into two-dimension data to be clustered into four bins. Even though, UMAP clusters are better at preserving density and distance than other clustering methods like t-SNE when used for dimensionality reduction, it is generally not advised to run clustering solely on the results of UMAP without the use of any cluster verification methods. Two-dimension data plots of UMAP dimensional reduction show overlapping points which are also difficult to identify in hard clustering density based methods.

To account for these short comings, Fuzzy c-means clustering method is deployed to capture the bins of RUL that need to schedule a maintenance. While other means of clustering like hard clustering based k-means and density-based clustering method like DBSCAN were considered, fuzzy c-means was selected due to its soft clustering capabilities and availability of the Fuzzy partition matrix. Fuzzy partition matrix helps make the results interpretable and opens up possibilities to inspect belongingness of the cycle to a given maintenance bin. Unlike k-means, fuzzy c-means also handles clusters of arbitrary shapes and does not suffer from misclassification errors due to lack of high-density points that affects density-based methods.

The eighteen dimension data cases are compared with another data case where SHAP data is used to select the top five most important variables to explore the viability of this framework in using sensor constrained data for PHM applications with XAI based SHAP information obtained from a NN. These data cases are summarized in Table \ref{table:data_cases}. Finally, to study the reliability of such clustering methods, validation scores are calculated for all four major clustering methods in scikit-learn modules -- Rand Index, Mutual Information (MI) based scores, V-measure and its group of scores and finally Fowlkes-mallows score.

\begin{table}[ht]
\centering
\begin{tabular}{c|c|c|c}
\hline
Data Case  & Raw Data or SHAP values & SHAP Informed & Number of Dimensions \\

\hline
&&&\\
1 & Raw Data & No & 18 \\ 
2 & Raw Data &  Yes & 5 \\ 
3 & SHAP values &  No & 18 \\ 
4 & SHAP values &  Yes & 5 \\
\hline
\end{tabular}
\caption{Data cases for Input Dimensions}
\label{table:data_cases}
\end{table}

\section{Results}
\label{sec:3}
In this section, we look at the results obtained by applying methods on the NASA C-MAPSS ’08 FD-001 dataset. Results from neural net and SHAP analysis are first explored, followed by fuzzy c-means clustering results of unsupervised raw variables and XAI based semi-supervised data. This is followed by applying the same clustering methods on SHAP values and a final gathering of cluster validation scores.

\subsection{Neural Net and SHAP Results}
\begin{figure}[hbt!]
\centering
\includegraphics[width = 0.65\linewidth]{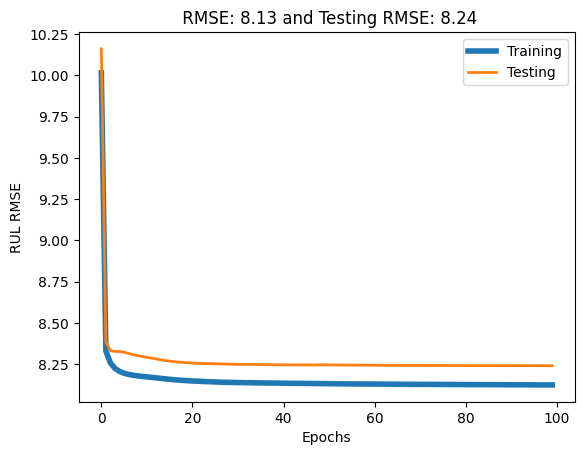}\caption{Testing vs Training RMSE for FD001 dataset}
\label{fig:RMSE_plot}
\end{figure}

\begin{figure}[hbt!]
\centering
\includegraphics[trim=0 0 0 20,clip, width = 0.65\linewidth]{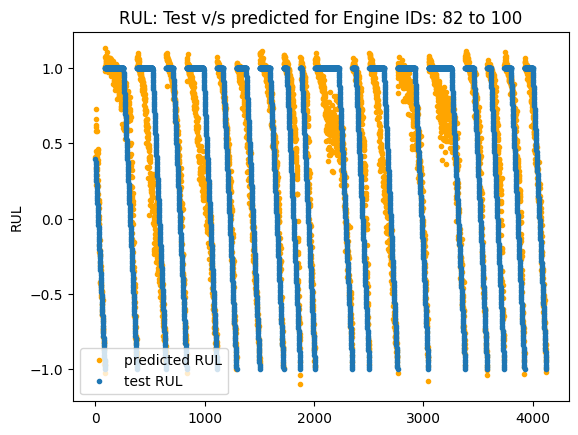}\caption{Predicted vs Testing RUL for FD001 dataset}
\label{fig:RUL_plot}
\end{figure}

A simple Artificial Neural Network was run for 100 epochs on the FD001 dataset as shown in Figure \ref{fig:RMSE_plot}. A few different variations on the epoch and hyper-parameters were tested before finalizing on this number of epochs. The RMSE value for testing data was $8.24$ and for training was $8.13$ While not the most accurate, these results were deemed acceptable for this usecase which was to prepare an input for eXplainable AI (XAI) module - SHAP in python. The resulting RUL predictions along with the piecewise \emph{ground truth} RUL is visualized in Figure \ref{fig:RUL_plot}. The abscissa indicates the testing row number for each engine row data.

\begin{figure}[hbt!]
\centering
\includegraphics[width = 0.65\linewidth]{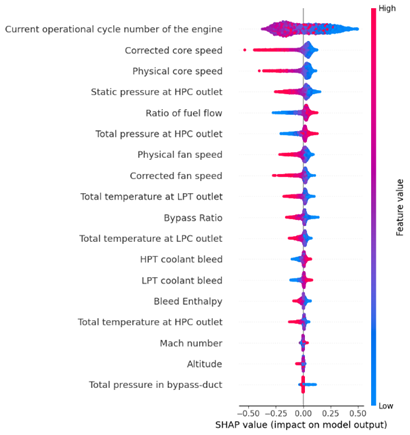}\caption{SHAP Summary plot for FD001 dataset}
\label{fig:SHAP_summary_plot}
\end{figure}

SHAP was run for all the testing row data along with its predicted RUL values. The resulting values were then visualized in a summary plot shown in Figure \ref{fig:SHAP_summary_plot}. This plot shows all the  SHAP values of the dimensions starting with one that most impacts the decision all the down with decreasing priority. The points of the SHAP values are also colored with value of the feature to emphasize if a given SHAP value is directly or indirectly proportional to the overall model output. For use in the next few sub sections, top five dimensions from this plot were recorded and exported. The SHAP values and raw variable data from each of these five variables will be used as a semi-supervised dataset and compared with unsupervised training results.

\subsection{Cluster Analysis Results}
UMAP was used to reduce dimensions of the dataset into a two dimensional scatter plots. We will discuss two different approaches to our data in this section and validate these clusters in the next section.
\subsubsection{Raw data Cluster Analysis}
\begin{figure}[hbt!]
\centering
\begin{tabular}{cc}
\subfloat[Raw variables - Ground truth]{\includegraphics[trim=0 0 175 15,clip, width = 0.40\linewidth]{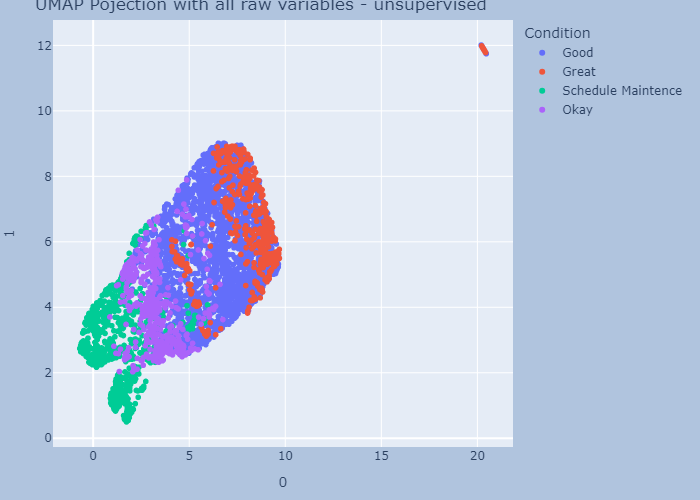}} &
\subfloat[Five raw variables - Semi-supervised with SHAP Summary plots]{\includegraphics[trim=0 0 175 15,clip,width = 0.40\linewidth]{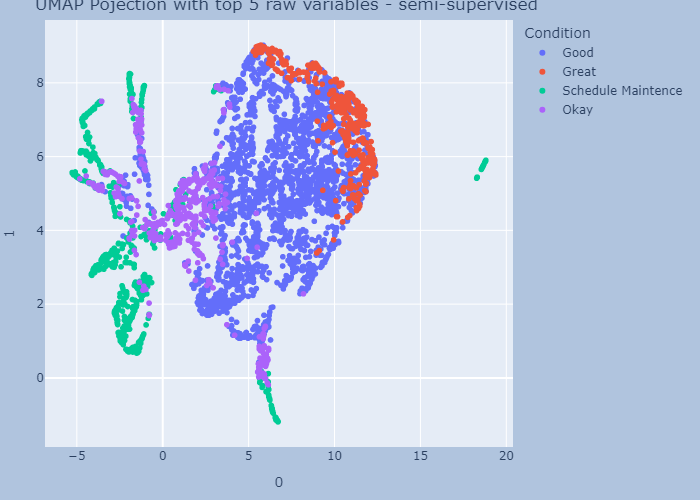}}\\

\subfloat[Unsupervised Fuzzy C-means cluster of all raw variables]{\includegraphics[trim=0 0 00 22,clip,width = 0.45\linewidth]{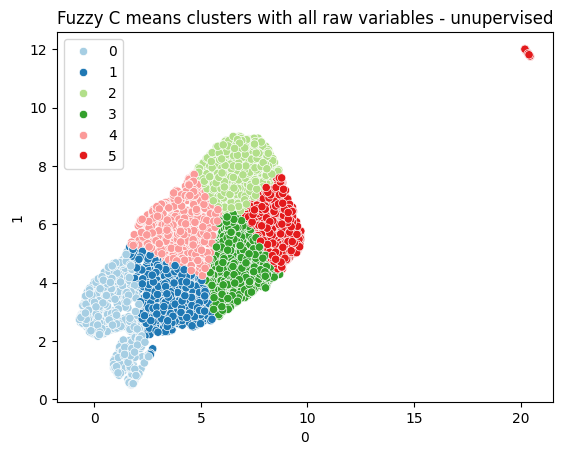}} &
\subfloat[Five raw variables picked using SHAP Summary plots]{\includegraphics[trim=0 0 20 22,clip,width = 0.45\linewidth]{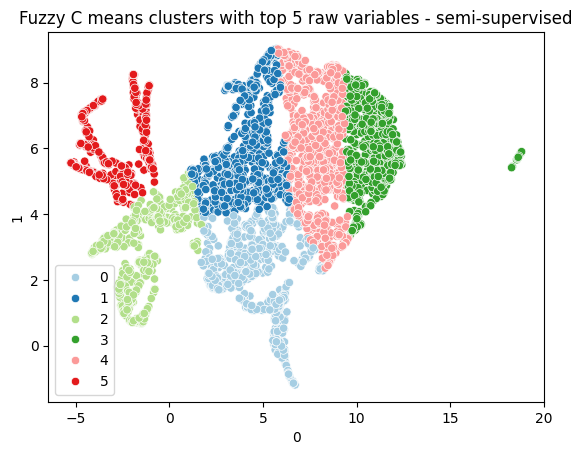}}\\

\end{tabular}
\caption{Raw data clusters - 2D UMAP dimension reduced}
\label{fig:raw_variables_clusters}
\end{figure}

Figures \ref{fig:raw_variables_clusters}a and \ref{fig:raw_variables_clusters}b show results from the dimensional reduction for unsupervised nineteen dimensions and semi-supervised 5 dimensions, respectively. These reduced 2D clusters are colored according to our afore mentioned \emph{ground truth} maintenance times predicted by the ANN and are listed in Table \ref{table:cluster_colors}. A simple glimpse highlights how complex the data clusters are and further highlights the needs for such frameworks. The most critical one is \emph{Schedule} bin which is highlighted in green and have a RUL of below fifty. For both these respective data cases, a six-cluster fuzzy c-means algorithm was run with a three value of degree of fuzziness. The obtained clusters are shown in Figures \ref{fig:raw_variables_clusters}c and \ref{fig:raw_variables_clusters}d for unsupervised and semi-supervised cases, respectively. A good majority of the green section for unsupervised case are captured in clusters zero, four, and three with some over flowing into cluster number one. For semi-supervised case, the fuzzy c-means captures \emph{Schedule} bin in five, two, and zero clusters.

\subsubsection{SHAP Clusters Analysis}

\begin{figure}[hbt!]
\centering
\begin{tabular}{cc}
\subfloat[All SHAP values]{\includegraphics[trim=0 0 175 15,clip,width = 0.40\linewidth]{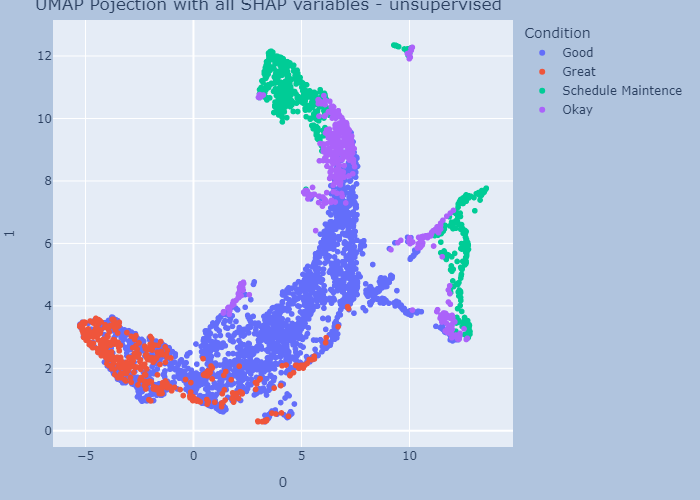}} &
\subfloat[Five raw variables - Semi-supervised with SHAP Summary plots]{\includegraphics[trim=0 0 175 15,clip,width = 0.40\linewidth]{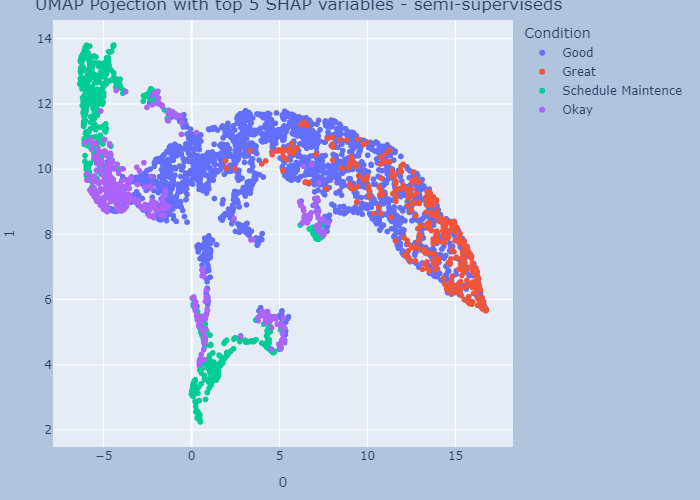}}\\

\subfloat[Unsupervised Fuzzy C-means cluster of all SHAP values]{\includegraphics[trim=0 0 00 22,clip,width = 0.45\linewidth]{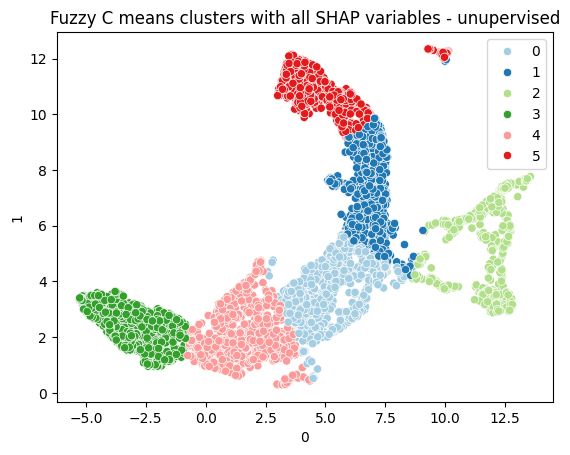}} &
\subfloat[Five SHAP values picked using SHAP Summary plots]{\includegraphics[trim=20 0 50 22,clip,width = 0.45\linewidth]{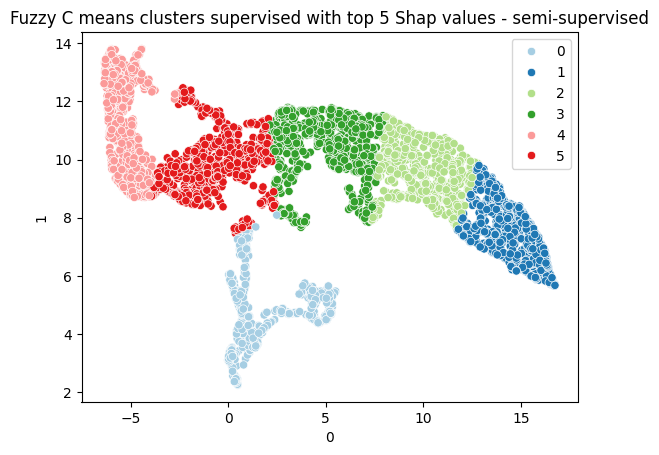}}\\

\end{tabular}
\caption{SHAP data clusters -- 2D UMAP dimension reduced}
\label{fig:SHAP_clusters}
\end{figure}

Similarly variable selection approach was taken for Shapley values obtained from XAI framework SHAP. All nineteen dimensions of Shapley values were reduced into a 2D data and colored according to their row data assigned RUL value color. Figure \ref{fig:SHAP_clusters}a and b show these plots for unsupervised shapely values and SHAP informed semi-supervised data, respectively. The fuzzy c-means cluster for unsupervised Shapley values that indicate the RUL row in the \emph{Schedule} domain are two, and 5 as shown in Figure \ref{fig:SHAP_clusters}c. Conversely, it required more than two clusters to capture the same number of \emph{Schedule} bins in semi-supervised scatter plot with fuzzy c-mean algorithm requiring cluster two, three, zero, and four for the same task.

\begin{table}[ht]
\centering
\begin{tabular}{c|c}
\hline
Color  & Maintenance Bins\\
\hline
\\
Red & Great \\ 
Blue & Good \\ 
Purple & Okay \\ 
Green & Schedule \\
\hline
\end{tabular}
\caption{RUL bins by color}
\label{table:cluster_colors}
\end{table}

\subsection{Clustering Validation Results}

\begin{figure}[hbt!]
\centering
\includegraphics[width = 0.60\linewidth]{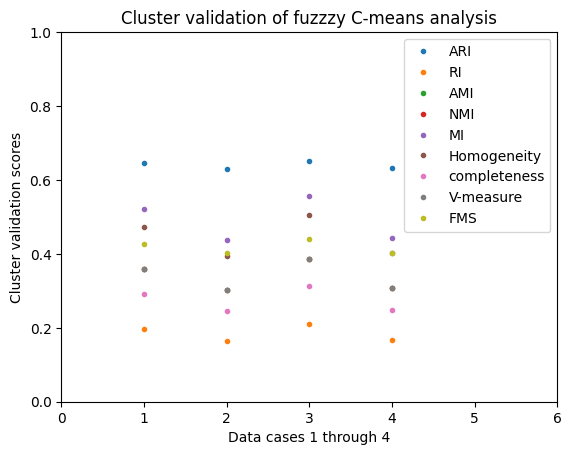}\caption{Data case vs Cluster validation Metrics}
\label{fig:clustering_validation}
\end{figure}

While visual inspection is a great tool and helps with basic understanding of similarity between clusters without the use of  any complex tools, a quantitative approach is need to identify the accuracy of clustering analysis and all four major available methods for datasets with accessible ground truth were used in this validation. The results for all four data cases are plotted in Figure \ref{fig:clustering_validation} and their respective numerical values expressed in Table \ref{table:clus_valid_table}. 

While clustering parameters can be further optimized to improve these metrics, the focus of this framework is primarily on how these clustering metrics vary across our two dimensions -- with raw and SHAP value data, and with five vs eighteen sensors along with the implications of those dimensions on RUL bins for sensor constrained applications.
All cluster validation metrics show improved performance when using SHAP values instead of Raw data. When using all eighteen sensor inputs, the difference in metrics for using SHAP values improves by a maximum of 0.03 for Mutual Index scores, Homogeneity, and V-measure.
This effect is less profound in sensor constrained applications with five sensors where a maximum difference of 0.01 is observed for SHAP values clusters in Mutual Index and Homogeneity metrics. All other metrics remain constant until the second significant digit. This implies that for sensor constrained data cases, we can simply use the maintenance bins to identify scheduling times from five sensors of raw data clusters instead of calculating clusters using SHAP values. Further work needs to be carried out in this regard to explore the number of sensors with an aim to obtain further reduction in sensor numbers.

\begin{table}[ht]
\centering
\begin{tabular}{c|c|c|c|c}
\hline
& Raw data & Sensor constrained  & SHAP values & Sensor constrained  \\
Clustering Validation   & & raw data & & SHAP values \\
metric &  \emph{18 Sensors} &  \emph{5 Sensors} &  \emph{18 Sensors} &  \emph{5 Sensors} \\
 & (Case 1) & (Case 2) & (Case 3) & (Case 4) \\
\hline
&&&&\\

ARI & 0.6452 & 0.6301 & 0.6511 & 0.6324 \\ 
RI & 0.1957 & 0.1634 & 0.2102 & 0.1670 \\ 
AMI & 0.3590 & 0.3026 & 0.3858 & 0.3065 \\ 
NMI & 0.3598 & 0.3035 & 0.3866 & 0.3074 \\
MI & 0.5204 & 0.4363 & 0.5571 & 0.4442 \\
Homogeneity & 0.4715 & 0.3953 & 0.5048 & 0.4025 \\
Completeness & 0.2909 & 0.2463 & 0.3132 & 0.2486 \\
V-measure & 0.3598 & 0.3035 & 0.3866 & 0.3074 \\
FMS & 0.4269 & 0.4022 & 0.4403 & 0.4031 \\
\hline
\end{tabular}
\caption{Cluster Validation Metrics and their numerical values}
\label{table:clus_valid_table}
\end{table}


\pagebreak
\section{Conclusion}
\label{sec:5}

In aircraft maintenance applications, it is generally seen if one can reduce monitoring and sensor based costs for an engine that is towards the end of its planned life cycle. A semi-supervised fault diagnosis approach is presented to address the problem of how relevant a given sensor is for a given decision and maintenance planning. The framework is applied to a NASA C-MAPSS ’08 dataset with four different data cases. The resulting output used 70\% less data while offering a similar cluster quality and decision-making abilities. For the same amount of data, using the framework produced higher quality of clusters and opens up further design space to optimize. The contributions in this work have practical implications in prognostics and health management of current jet engines along with other associated PHM allied industries like finance and healthcare.

%
%
%


\end{document}